\documentclass[journal]{IEEEtran}

\usepackage{cite}
\usepackage{graphicx}
\usepackage[caption=false,font=footnotesize]{subfig}
\usepackage{amsmath}
\usepackage{url}
\usepackage{amssymb}

\begin{document}

\title{JAX-Based Batched AC Power Flow for GPU Acceleration and AI Ecosystem Integration}

\author{Yihong Zhou,~\IEEEmembership{Member,~IEEE}, Dylan Cope, Jakob Foerster, and Thomas Morstyn,~\IEEEmembership{Senior Member,~IEEE}%
\vspace{-1mm}
\thanks{Department of Engineering Science, University of Oxford, U.K.}%
\thanks{Corresponding author: Yihong Zhou (e-mail: yihong.zhou@eng.ox.ac.uk).}}

\markboth{IEEE Power Engineering Letters}%
{}

\maketitle

\begin{abstract}
Coordinating growing grid flexibility under uncertainty is becoming increasingly important for efficient and reliable power-system operation. A core computational requirement is the efficient large-scale batched evaluation of AC power flow across candidate operating actions and uncertainty scenarios. Previous work has explored GPU-based batched power-flow evaluation, but has largely relied on hand-written C or CUDA code, creating barriers to customisation, efficient kernel optimisation, and long-term maintenance. JAX is a Python-based framework that enables efficient accelerator execution while keeping implementations in Python. This letter therefore proposes a JAX-based batched AC power-flow solver that uses current JAX functionality to implement Newton--Raphson for transmission networks and Z-Bus power flow for three-phase unbalanced distribution networks, achieving more than $10\times$ speed-ups relative to \texttt{pandapower} and \texttt{OpenDSS}. In addition, JAX integrates seamlessly with the broader JAX-based AI ecosystem, making it straightforward to embed power-flow evaluation within AI methods for future larger-scale and more complex power-system operation.
\end{abstract}


\vspace{-3mm}
\section{Introduction}

\IEEEPARstart{P}{ower} systems are facing growing congestion and uncertainty as variable renewable generation increases and transport and heating become electrified. To reduce time-consuming and capital-intensive network infrastructure upgrades, it is increasingly important to deploy advanced decision-making methods that coordinate flexible resources across the network, including the millions of grid-edge devices such as heat pumps and electric vehicles being integrated \cite{nesonto}. For reliable system operation, a core component in such coordination is the large-scale batched evaluation of accurate AC power flow across candidate actions or uncertainty scenarios.
% , as well as network-topology changes such as bus splitting/merging and line dis-/reconnection that are being developed in industry

Accelerators such as graphics processing units (GPUs) are promising for batched power-flow calculations because of their high floating-point throughput and memory bandwidth. Prior work has developed GPU-based methods for accelerated batched AC power-flow computation \cite{zhou2017gpu, zhou2017gpu2, su2020gpu}. However, these methods are largely programmed in low-level frameworks such as C/CUDA, creating a high barrier to customisation, further development, and long-term maintenance.
 
JAX is attractive because it allows the solver to be expressed in Python/NumPy-style code while supporting efficient program transformations \cite{jax2018}, which stage computations to XLA for compilation into optimised accelerator code and can, in many cases, achieve performance close to hand-written C or CUDA code \cite{jax2018}. Although recent work has also explored Julia for batched GPU power flow \cite{alla2025accelerating}, JAX offers an additional practical advantage through its seamless integration with the broader JAX-based AI ecosystem, including reinforcement learning (RL), imitation/supervised learning, and other learning-based models such as large language models (LLMs). This avoids the additional runtime overhead that can arise when coupling a power-flow solver implemented outside a specific AI stack, and allows the entire learning pipeline to reside on the GPU---for instance, JAX-based end-to-end multi-agent RL frameworks have achieved training speedups of up to 12,500$\times$ over conventional approaches \cite{rutherford2024jaxmarl}.
Such efficient AI compatibility is becoming increasingly important for future power-system operation with millions of controllable devices, where conventional coordination pipelines may become difficult to scale and learning-based methods play a growing role.

This letter therefore proposes JAX-based batched AC power flow as a practical software layer. Although JAX’s built-in sparse support remains limited and largely experimental compared with mature CPU sparse ecosystems, we show that current JAX functionality already supports efficient batched power-flow solvers in two representative settings: Newton--Raphson for transmission networks and Z-Bus power flow (current injection) for three-phase unbalanced distribution networks.

\vspace{-3mm}
\section{Methods}

% The unifying design principle is to express both solvers as sparse iterative maps that JAX can batch and compile while keeping the implementation close to standard formulations. This preserves a Python workflow that researchers can still customize while producing solver components that can be embedded directly into JAX-based AI, rollout, optimisation, or learning loops. For transmission systems, we use sparse Newton--Raphson with Jacobian-vector-product generalised minimal residual (GMRES) iterations and a structured preconditioner. For distribution systems, we use a three-phase Z-bus fixed-point formulation with batched wye and delta load injections.

\begin{figure*}[!t]
\centering
\vspace{-6mm}
\subfloat[]{\includegraphics[width=0.5\textwidth]{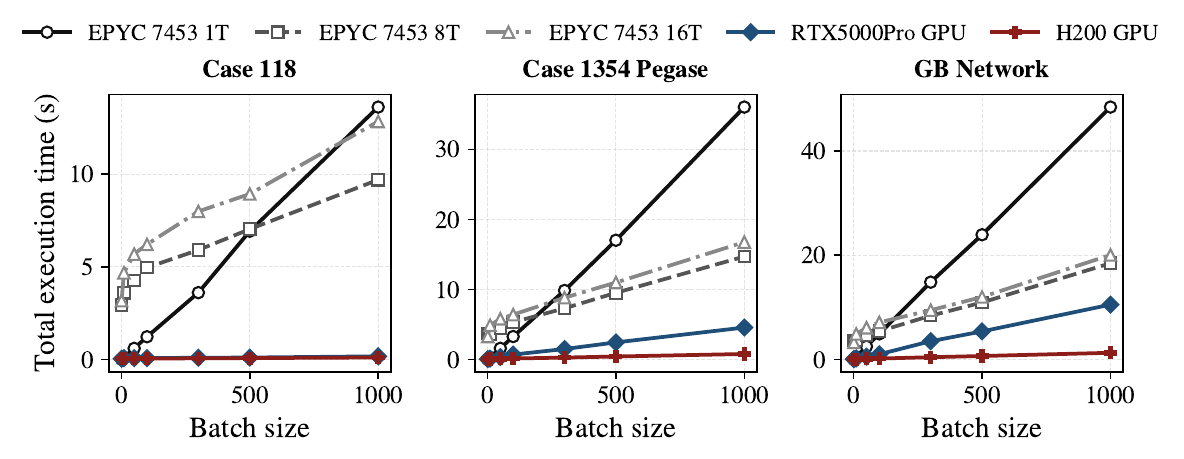}\label{fig:tx-total}}
\hfill
\subfloat[]{\includegraphics[width=0.5\textwidth]{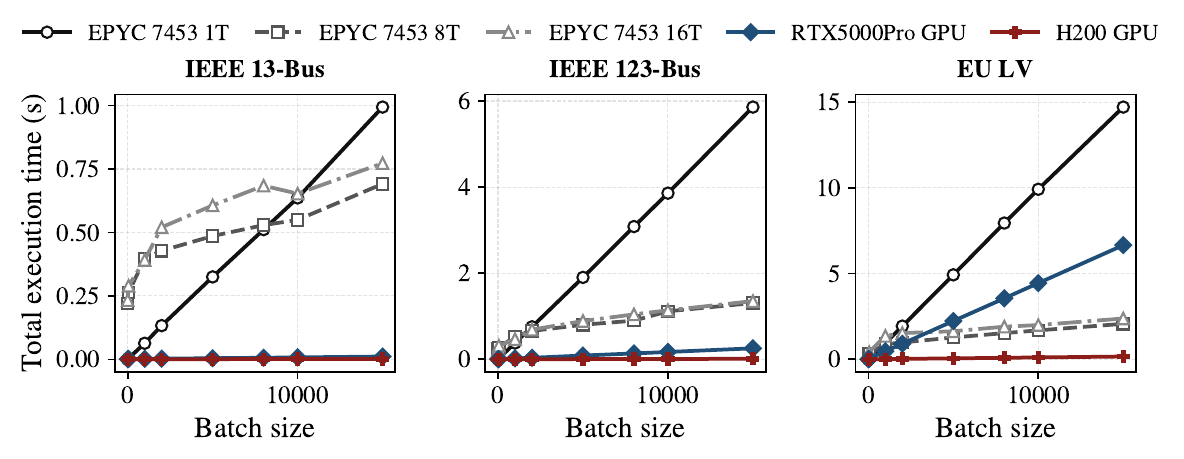}\label{fig:dist-total}}\\[-1.8ex]
\subfloat[]{\includegraphics[width=0.495\textwidth]{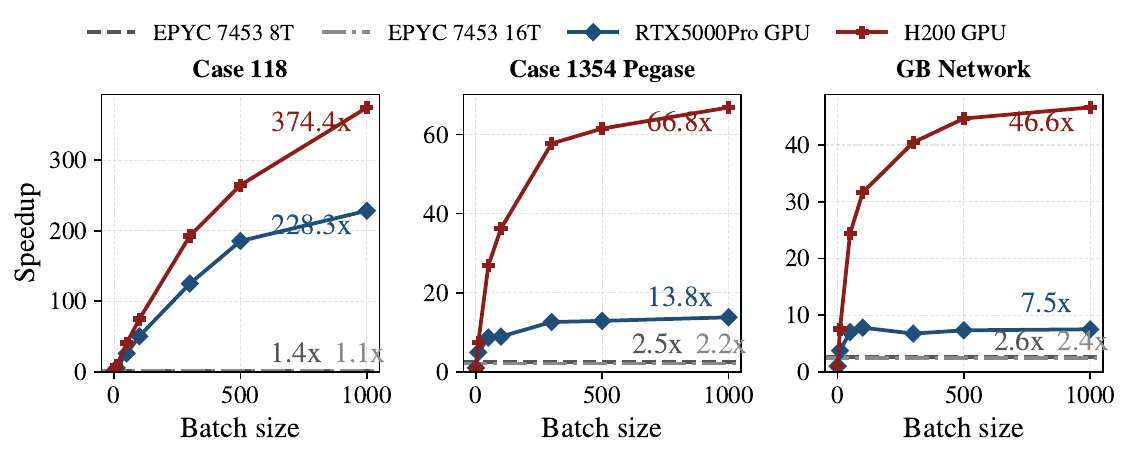}\label{fig:tx-inst}}
\hfill
\subfloat[]{\includegraphics[width=0.495\textwidth]{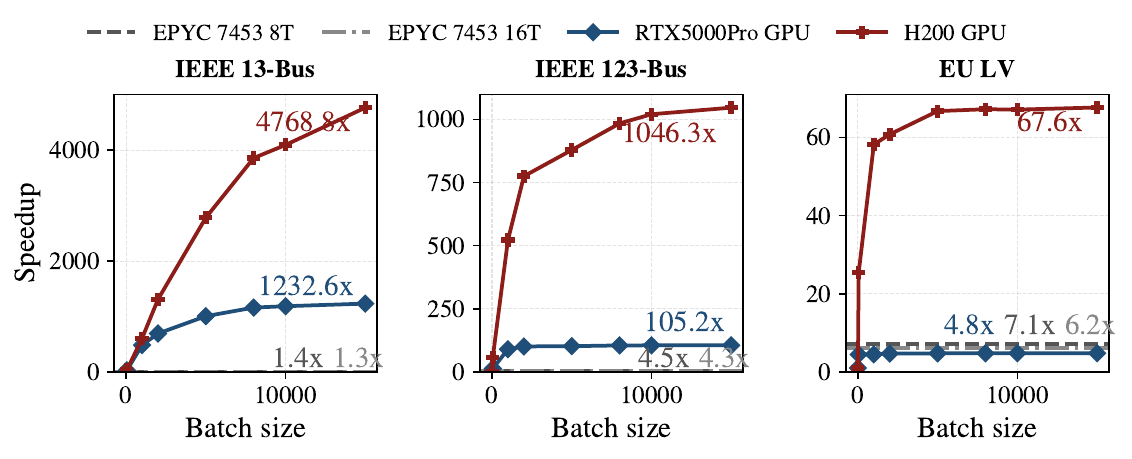}\label{fig:dist-inst}}
\caption{(a) Total computing time of batched AC power flow using the Newton--Raphson method in the transmission cases. \texttt{GBnetwork} has 2224 nodes in total. (b) Total computing time of batched AC power flow using the Z-Bus method in the three-phase unbalanced distribution cases. \texttt{EULV} has 2724 connection phases in total. (c) and (d) show the corresponding speed-up ratios relative to the single-threaded CPU solve (or the single-scenario GPU case), where the speed-up for all CPU and GPU implementations eventually plateaus, indicating that the hardware resources are fully utilised at these batch scales.}
\vspace{-1mm}
\label{fig:bench}
\end{figure*}

\subsection{Transmission: Newton--Raphson Power Flow in JAX}

% For transmission systems, we use a sparse Newton--Raphson algorithm with the generalised minimal residual method (GMRES) as the iterative Krylov linear solver, because batched sparse LU solvers are less well supported in JAX. 
We follow a standard AC power-flow formulation and solve the
resulting nonlinear equations using a sparse Newton--Raphson method.
In our implementation, GMRES is used as the Krylov linear solver for the Newton update step,
as efficient batched sparse direct solvers are currently limited in JAX.
Let $P_{b,\Theta}$ denote the active-injection vector over PV and PQ buses for scenario $b$, and let $Q_{b,\mathcal Q}$ denote the reactive-injection vector over PQ buses. We further let $\mathcal{PV}$ and $\mathcal{PQ}$ denote the PV- and PQ-bus index sets, and define $\Theta:=\mathcal{PV}\cup\mathcal{PQ}$ (the angle-state block) and $\mathcal{Q}:=\mathcal{PQ}$ (the PQ-voltage block). The JAX solver uses a standard AC Newton--Raphson formulation over the free variables (voltage angles $\theta_{b,\Theta}$ and magnitudes $V_{b,\mathcal{Q}}$)
\begin{equation}
x_b = \begin{bmatrix}\theta_{b,\Theta}^\top & V_{b,\mathcal{Q}}^\top\end{bmatrix}^{\!\top},
\end{equation}
with mismatch map
\begin{equation}
F_b(x_b)=
\begin{bmatrix}
P_{\mathrm{calc},\Theta}(x_b)-P_{b,\Theta}\\
Q_{\mathrm{calc},\mathcal{Q}}(x_b)-Q_{b,\mathcal{Q}}
\end{bmatrix}.
\end{equation}
At Newton iteration $k$, the state update solves
\begin{equation}
J_b^k \Delta x_b = -F_b(x_b^k), \qquad x_b^{k+1}=x_b^k+\Delta x_b,
\end{equation}
where $J_b^k$ is the Jacobian.

The key implementation choice is to solve the above Jacobian linear system \emph{iteratively}, rather than explicitly factorising the Jacobian. The bus-admittance matrix $Y_{\mathrm{bus}}$ is converted to coordinate (COO) format, split into real and imaginary parts, and stored as JAX sparse \texttt{BCOO} tensors. Sparse products are evaluated with \texttt{bcoo\_dot\_general}. At each Newton iteration, \texttt{jax.linearize} provides a Jacobian-vector-product operator for the mismatch map, and GMRES computes the correction step.

Effective preconditioning is important for Krylov solvers because it improves the numerical properties of the linear operator, which typically leads to faster and more robust GMRES convergence. An ideal preconditioner should be inexpensive to apply while approximating the inverse of the coefficient matrix ($J_b^k$ in our case). To design this preconditioner, we first approximate $J_b^k$ with a sparse lower-triangular operator built from the standard polar-form Jacobian partition
\begin{equation}
J_b^k=
\begin{bmatrix}
H_b & N_b\\
M_b & L_b
\end{bmatrix},
\end{equation}
where $H_b=\partial P_{\mathrm{calc},\Theta}/\partial \theta_{b,\Theta}$, $N_b=\partial P_{\mathrm{calc},\Theta}/\partial V_{b,\mathcal{Q}}$, $M_b=\partial Q_{\mathrm{calc},\mathcal{Q}}/\partial \theta_{b,\Theta}$, and $L_b=\partial Q_{\mathrm{calc},\mathcal{Q}}/\partial V_{b,\mathcal{Q}}$. In the fast-decoupled approximation \cite{fast_decoupled_load_flow}, the dominant terms satisfy $H_b \approx \mathrm{diag}(V_{b,\Theta}^k) B'$ and $L_b \approx \mathrm{diag}(V_{b,\mathcal Q}^k) B''$, with $B'$ and $B''$ being the corresponding principal submatrices of $-\mathrm{Im}(Y_{\mathrm{bus}})$, namely $B' := -\mathrm{Im}(Y_{\mathrm{bus}})_{\Theta,\Theta}$ and $B'' := -\mathrm{Im}(Y_{\mathrm{bus}})_{\mathcal{Q},\mathcal{Q}}$.

To retain the leading cross coupling, we approximate $M_b$ with a sparse conductance-driven surrogate. Let $\theta_{ij}=\theta_i-\theta_j$, and let $G_{ij}$ and $B_{ij}$ denote the conductance and susceptance entries of $Y_{\mathrm{bus}}$. The elements in $M_b$ are
\begin{align}
[M_b]_{i,j} &
=
-|V_i||V_j|\left(G_{ij}\cos \theta_{ij} + B_{ij}\sin \theta_{ij}\right), \  & i \neq j,\\
[M_b]_{i,i} &
=
P_i-G_{ii}|V_i|^2, \  & i = j.
\end{align}
Under the usual fast-decoupled assumptions of small angle differences and near-nominal voltage magnitudes, $\sin \theta_{ij}\approx 0$ and $\cos \theta_{ij}\approx 1$, so the off-diagonal part is approximately conductance-driven. Our implementation therefore keeps a sparse conductance-based surrogate while building the left preconditioner $(\mathcal{M}_b^k)^{-1}$ with
\begin{equation}
\mathcal{M}_b^k :=
\begin{bmatrix}
\mathrm{diag}(V_{b,\Theta}^k) B' & 0\\
G_{\mathcal{Q},\Theta} & \mathrm{diag}(V_{b,\mathcal Q}^k) B''
\end{bmatrix},
\end{equation}
where $G_{\mathcal{Q},\Theta}$ is the corresponding sub-block of $-\mathrm{Re}(Y_{\mathrm{bus}})$. Applying $(\mathcal{M}_b^k)^{-1}$ only requires forward block substitution. For a residual $r_b=[r_{b,\Theta}^\top, r_{b,\mathcal{Q}}^\top]^\top$ in GMRES, applying the preconditioner gives $z_b := [z_{b,\Theta}^\top, z_{b,\mathcal{Q}}^\top]^\top = (\mathcal{M}_b^k)^{-1}\, r_b$, which can be computed as:
\begin{align}
z_{b,\Theta} &= (B')^{-1}\mathrm{diag}(V_{b,\Theta}^k)^{-1} r_{b, \Theta},\\
z_{b,\mathcal{Q}} &= (B'')^{-1}\mathrm{diag}(V_{b,\mathcal Q}^k)^{-1} \left(r_{b, \mathcal{Q}} - G_{\mathcal{Q},\Theta}\, z_{b,\Theta} \right).
\end{align}
To stabilise the precomputed inverses, we use $B' + \varepsilon I$ and $B'' + \varepsilon I$ with a small diagonal regularisation $\varepsilon$. The scenario dimension is mapped by \texttt{vmap} and compiled with \texttt{jit}, removing Python-side loop overhead from the batched Newton pipeline.

\vspace{-3mm}
\subsection{Distribution: Z-Bus Power Flow in JAX}

For distribution studies, we use a three-phase Z-Bus formulation based on the current-injection method with a precomputed reduced bus-impedance matrix \cite{alla2025accelerating}. Ref. \cite{alla2025accelerating} also notes that additional GPU speed-up can be achieved by exploiting the sparsity of the admittance matrix, which we leave for future work. After separating slack and non-slack phases, the fixed-point map for scenario $b$ is
\begin{equation}
v_b^{k+1}=Z\,i\!\left(v_b^k,S_{b}^{\mathrm{wye}},S_{b}^{\mathrm{del}}\right)+v_0,
\end{equation}
where $Z$ is the reduced bus-impedance matrix, $v_0$ is the no-load voltage term, and $i(\cdot)$ denotes the nonlinear current-injection map induced by the wye and delta loads. Wye currents are computed as $-(S_b^{\mathrm{wye}}/v_b^k)^{*}$, whereas delta currents are formed from line-to-line voltages and then mapped back to phase currents. The JAX implementation runs this fixed-point iteration inside a compiled loop until the change in the sum of voltage magnitudes is sufficiently small or the maximum iteration number is reached. The scenario dimension is again mapped by \texttt{vmap} and compiled with \texttt{jit} for efficient batched GPU execution.

\vspace{-3mm}
\section{Case Studies}
We compare our JAX solvers with \texttt{pandapower} for transmission networks and \texttt{OpenDSS} for three-phase unbalanced distribution networks, both of which are widely used software packages for power-system analysis. Scenario batches are generated from $\pm 20$\% random load multipliers and converted into perturbed injection vectors. For transmission, the CPU baseline runs \texttt{pandapower.runpp} with fixed Newton--Raphson settings, flat start, voltage-angle calculation enabled, and matched tolerances, using \texttt{joblib} parallelism with 1, 8, and 16 threads. The test network cases are \texttt{case118}, \texttt{case1354pegase}, and the 2224-node \texttt{GBnetwork}. The maximum absolute difference between the JAX and \texttt{pandapower} solutions is below $10^{-10}$.

For distribution, the \texttt{OpenDSS} solver is adjusted to align with the JAX assumptions as closely as possible: snapshot mode, all control actions disabled, a constant-power load model, and repeated no-control solves. The test networks are \texttt{IEEE13}, \texttt{IEEE123}, and \texttt{EULV}, the last of which has 2724 phase connections in total. Reported JAX timings exclude compilation warm-up, and \texttt{OpenDSS} timings reuse a warmed-up engine to emphasise steady-state throughput. The maximum absolute voltage-magnitude differences between the JAX and \texttt{OpenDSS} results are $3\times10^{-4}$ p.u. for \texttt{IEEE13}, $2\times10^{-5}$ p.u. for \texttt{IEEE123}, and $7\times10^{-6}$ p.u. for \texttt{EULV}; these differences are likely attributable to differences in network-parameter handling. We have also validated the JAX implementation against a separate Numpy-based Z-Bus load-flow solver in OPEN \cite{morstyn2020open}.

All experiments use a shared Python environment with Python 3.11.14, JAX 0.9.0.1, \texttt{pandapower} 3.4.0, \texttt{py\_dss\_interface} 2.3.0, and \texttt{joblib} 1.5.3. CPU baselines are obtained on an AMD EPYC 7453 28-Core Processor with 200~GB RAM, while GPU results are obtained on an NVIDIA RTX (PRO) 5000 Blackwell 48~GB and an NVIDIA H200 141~GB. Double-precision floating-point arithmetic is enabled for GPU implementation, which means the performance of the non-data-centre-grade RTX5000 GPU can be limited.

Figs.~\ref{fig:tx-total} and \ref{fig:dist-total} show the overall computational benefits of our JAX-based batched AC power flow. For relatively small networks, both the RTX5000 and the H200 GPUs exhibit nearly constant runtime as the batch size increases. As the network size increases to the largest case, the runtime on the RTX5000 grows at a rate comparable to that of the multi-threaded CPU benchmark and can even exceed that of the CPU benchmark for the \texttt{EULV} distribution network. This is because the RTX5000 is not a data-centre-grade GPU and has limited double-precision throughput. In contrast, for the data-centre-grade H200 GPU, total runtime remains nearly constant across all cases, yielding speed-ups of $14.3\times$ and $12.9\times$ relative to the fastest CPU benchmark for the largest 2224-node \texttt{GBnetwork} at batch size 1,000 and the 2724-node-phase \texttt{EULV} network at batch size 15{,}000, respectively.

Figs.~\ref{fig:tx-inst} and \ref{fig:dist-inst} further show the benefits of GPU parallelism compared with CPU multi-threading. It can be seen that 16-thread CPU execution is sometimes slower than 8-thread CPU execution because of limited memory bandwidth and the overhead of Python-side thread management. Also, whereas CPU multithreading delivers at most a 7$\times$ speed-up relative to sequential single-scenario CPU execution, GPU parallelism achieves up to a 4{,}700$\times$ speed-up compared to single-scenario GPU execution.

\vspace{-3mm}
\section{Discussion and Conclusion}

To support large-scale power system operation, this letter presents JAX-based batched AC power-flow solvers, in which conventional algorithms are recast to fully leverage native JAX capabilities. The proposed approach has two main advantages. First, \texttt{jit} and \texttt{vmap} support concise implementation and efficient accelerator execution of batched power-flow computations, delivering speed-ups of $14.3\times$ for Newton--Raphson on the 2224-node GB transmission network and $12.9\times$ for Z-Bus power flow on the 2724-node-phase three-phase unbalanced distribution network. Compared with a C/CUDA-based code base, the resulting implementation is more readable and easier to customise, maintain, and extend. Second, remaining within JAX enables seamless integration with the broader JAX-based AI ecosystem, which may support future larger-scale and more complex power-system operation.

At present, JAX's sparse support is less mature than conventional sparse linear-algebra stacks, which is why we use an indirect GMRES solve in Newton--Raphson and the \texttt{BCOO} format in the present implementation. Nevertheless, we have shown that current JAX can already deliver more than a $10\times$ computational advantage over CPUs. Looking ahead, customised interfaces via the JAX foreign-function interface (FFI) could also be used to integrate third-party sparse engines into JAX, such as NVIDIA cuDSS, which provides GPU sparse direct factorisation and solve routines, including batched modes.

\bibliographystyle{IEEEtran}

% \vspace{-2mm}
\bibliography{ref}

\end{document}